\documentstyle[12pt]{article}

\font\tenrm=cmr10

\font\elevenbf=cmbx10 scaled\magstep 1
 1
 1

\def\ie{{\em i.e.}}
\def\eg{{\em e.g.}}

\def\beq{\begin{equation}}
\def\eeq{\end{equation}}

\catcode`\@=11 
\def\coeff#1#2{{\textstyle{#1\over #2}}}

\def\vev#1{\left\langle #1\right\rangle}
\def\lsim{\mathrel{\mathpalette\@versim<}}
\def\gsim{\mathrel{\mathpalette\@versim>}}
\def\@versim#1#2{\vcenter{\offinterlineskip
    \ialign{$\m@th#1\hfil##\hfil$\crcr#2\crcr\sim\crcr } }}

\def\JL{J. L. Lopez}
\def\DVN{D. V. Nanopoulos}

\def\r#1{$\elevenbf#1$}
\def\rb#1{$\elevenbf\overline{#1}$}

\def\t1{{\tilde 1}}
\def\ov{\overline}

\def\mpt{p\hskip-5.5pt/\hskip2pt}

\def\GeV{\,{\rm GeV}}
\def\TeV{\,{\rm TeV}}

\def\wt{\widetilde}

\def\to{\rightarrow}
\def\pb{\,{\rm pb}}
\def\ipb{\,{\rm pb}^{-1}}
\def\NPB#1#2#3{Nucl. Phys. B {\elevenbf#1} (19#2) #3}
\def\PLB#1#2#3{Phys. Lett. B {\elevenbf#1} (19#2) #3}

\def\PRD#1#2#3{Phys. Rev. D {\elevenbf#1} (19#2) #3}
\def\PRL#1#2#3{Phys. Rev. Lett. {\elevenbf#1} (19#2) #3}
\def\PRT#1#2#3{Phys. Rep. {\elevenbf#1} (19#2) #3}
\def\MODA#1#2#3{Mod. Phys. Lett. A {\elevenbf#1} (19#2) #3}

\def\TAMU#1{Texas A \& M University preprint CTP-TAMU-#1}

\begin{document}
\begin{flushright}
{CERN-TH.6926/93}\\
{CTP-TAMU-33/93}\\
{ACT-12/93}\\
\end{flushright}
%


\begin{center}
\vglue 0.3cm
{\Large\bf SU(5) x U(1): A String Paradigm of a TOE and \\}
\vspace{0.2cm}
{\Large\bf its Experimental Consequences}
\footnote{To appear in the Proceedings of the INFN Eloisatron Project 26th
Workshop ``From Superstrings to Supergravity", Erice, Italy,
Dec. 5-12, 1992; D. Nanopoulos and A. Zichichi conference speakers.}\\
\vglue 0.7cm
{JORGE L. LOPEZ$^{(a),(b)}$, D. V. NANOPOULOS$^{(a),(b),(c)}$, and A.
ZICHICHI$^{(d)}$\\}
\vglue 0.4cm
{\em $^{(a)}$Center for Theoretical Physics, Department of Physics, Texas A\&M
University\\}
{\em College Station, TX 77843--4242, USA\\}
{\em $^{(b)}$Astroparticle Physics Group, Houston Advanced Research Center
(HARC)\\}
{\em The Woodlands, TX 77381, USA\\}
{\em $^{(c)}$CERN, Theory Division, 1211 Geneva 23, Switzerland\\}
{\em $^{(d)}$CERN, 1211 Geneva 23, Switzerland\\}
\baselineskip=12pt

\vglue 0.6cm
{\tenrm ABSTRACT}
 \end{center}

{\rightskip=3pc
 \leftskip=3pc
\xpt\baselineskip=12pt
\noindent
We present a string-inspired/derived supergravity model based on the flipped
$SU(5)\times U(1)$ structure supplemented by a minimal set of additional matter
representations such that unification occurs at the string scale
($\sim10^{18}\GeV$). This model is complemented by two string supersymmetry
breaking scenaria: the $SU(N,1)$ no-scale supergravity model and a
dilaton-induced supersymmetry breaking scenario. Both imply universal soft
supersymmetry breaking parameters: $m_0=0, A=0$ and
$m_0=\coeff{1}{\sqrt{3}}m_{1/2}, A=-m_{1/2}$ respectively. In either case the
models depend on only three parameters: $m_t$, $\tan\beta$, and $m_{\tilde g}$.
We present a comparative study of the sparticle and Higgs spectra of both
models and conclude that even though both can be partially probed at the
Tevatron, LEPII, and HERA, a larger fraction of the parameter space of the
no-scale model is actually accessible. In both cases there is a more
constrained version which allows to determine $\tan\beta$ in terms of
$m_t,m_{\tilde g}$. In the strict no-scale case we find that the value of $m_t$
determines the sign of $\mu$ ($\mu>0:\,m_t\lsim135\GeV$,
$\mu<0:\,m_t\gsim140\GeV$) and whether the lightest Higgs boson mass is above
or below $100\GeV$. In the more constrained version of the dilaton scenario,
$\tan\beta\approx1.4-1.6$ and $m_t\lsim155\GeV$, $61\GeV\lsim m_h\lsim91\GeV$
follow. Thus, continuing Tevatron top-quark searches and LEPI,II Higgs searches
could probe this restricted scenario completely.}
\begin{flushleft}
{CERN-TH.6926/93}\\
{CTP-TAMU-33/93}\\
{ACT-12/93}\\
June 1993
\end{flushleft}
\vfill\eject
\tableofcontents
\vfill\eject

\setcounter{page}{1}
\pagestyle{plain}

\section{Introduction}
\baselineskip=14pt
The ultimate unification of all particles and interactions has string theory as
the best candidate. If this theory were completely understood, we would be able
to show that string theory is either inconsistent with the low-energy world
or supported by experimental data. Since our present knowledge of string theory
is at best fragmented and certainly incomplete, it is important to consider
models which incorporate as many stringy ingredients as possible. The number of
such models is expected to be large, however, the basic ingredients that such
``string models" should incorporate fall into few categories: (i) gauge group
and matter representations which unify at
a calculable model-dependent string unification scale; (ii) a hidden sector
which becomes strongly interacting at an intermediate scale and triggers
supersymmetry breaking with vanishing vacuum energy and hierarchically small
soft superpersymmetry breaking parameters; (iii) acceptable high-energy
phenomenology, \eg, gauge symmetry breaking to the Standard Model (if needed),
not-too-rapid proton decay, decoupling of intermediate-mass-scale unobserved
matter states, etc.; (iv) radiative electroweak symmetry breaking; (v)
acceptable low-energy phenomenology, \eg, reproduce the observed spectrum of
quark and lepton masses and the quark mixing angles, sparticle
and Higgs masses not in conflict with present experimental bounds,
not-too-large neutralino cosmological relic density, etc.

All the above are to be understood as constraints on potentially realistic
string models. Since some of the above constraints can be independently
satisfied in specific models, the real power of a string model rests in the
successful satisfaction of all these constraints within a single model.

String model-building is at a state of development where large numbers of
models can be constructed using various techniques (so-called formulations)
\cite{models}. Such models provide a gauge group and associated set of matter
representations, as well as all interactions in the superpotential, the
K\"ahler potential, and the gauge kinetic function. The effective string
supergravity can then be worked out and thus all the above constraints can in
principle be enforced. In practice this approach has never been followed in its
entirety: sophisticated model-building techniques exist which can produce
models satisfying constraints (i), (iii), (iv) and part of (v); detailed
studies of supersymmetry breaking triggered by gaugino condensation
have been performed for generic hidden sectors; and extensive explorations of
the soft-supersymmetry breaking parameter space satisfying constraints (iii),
(iv), and (v) have been conducted.

In searching for good string model candidates, we are faced with two kinds of
choices to be made: the choice of the gauge and matter content of the model,
and the choice of the supersymmetry breaking mechanism. Fortunately, a string
theory theorem provides significant enlightenment regarding the first choice:
models whose gauge groups are constructed from level-one Kac-Moody algebras
do not allow adjoint or higher representations in their spectra \cite{ELN}.
This implies that the traditional GUT groups ($SU(5),SO(10),E_6$) are
excluded since the GUT symmetry would remain unbroken. Exceptions to this
theorem exist if one uses the technically complicated higher-level Kac-Moody
algebras \cite{Lewellen}, but these models are beset with constraints
\cite{ELN}. If one imposes the aesthetic constraint of unification of the
Standard Model non-abelian gauge couplings, then flipped $SU(5)\times U(1)$
\cite{Barr,revitalized,revamp,faspects} emerges as the prime candidate, as we
shortly discuss. String models without non-abelian unification, such as the
standard-like models of Refs. \cite{SMOrb,SMFFF}
and the Pati-Salam--like model of Ref. \cite{ALR} possess nonetheless gauge
coupling unification at the string scale, even though no larger structure is
revealed past this scale. However, the degree of phenomenological success
which some of these models enjoy, usually rests on some fortuitous set of
vanishing couplings which are best understood in terms of remnants of higher
symmetries.

Besides the very economic GUT symmetry breaking mechanism in flipped $SU(5)$
\cite{Barr,revitalized} -- which allows it to be in principle derivable from
superstring theory \cite{revamp} -- perhaps one of the more interesting
motivations for considering such a unified gauge group is the natural avoidance
of potentially dangerous dimension-five proton decay operators \cite{faspects}.
In Ref. \cite{LNZb} we constructed a supergravity model based on this gauge
group, which has the additional property of unifying at a scale $M_U={\cal
O}(10^{18})\GeV$, as expected to occur in string-derived versions of this model
\cite{Lacaze}. As such, this model constitutes a blueprint for string model
builders. In fact, in Ref. \cite{LNY} one such model was derived from string
and served as inspiration for the field theory model in Ref. \cite{LNZb}.
The string unification scale should be contrasted with the naive unification
scale, $M_U={\cal O}(10^{16}\GeV)$, obtained by running the Standard Model
particles and their superpartners to very high energies. This apparent
discrepancy of two orders of magnitude \cite{msu} creates a {\em gap} which
needs to be bridged somehow in string models. It has been shown \cite{price}
that the simplest solution to this problem is the introduction in the spectrum
of heavy vector-like particles with Standard Model quantum numbers. The minimal
such choice \cite{sism}, a quark doublet pair $Q,\bar Q$ and a $1/3$--charge
quark singlet pair $D,\bar D$, fit snugly inside a \r{10},\rb{10} pair of
flipped $SU(5)$ representations, beyond the usual
$3\cdot({\bf10}+\ov{\bf5}+{\bf1})$ of matter and \r{10},\rb{10} of Higgs.

In this model, gauge symmetry breaking occurs due to vacuum expectation values
(vevs) of the neutral components of the \r{10},\rb{10} Higgs representations,
which develop along flat directions of the scalar potential. There are two
known ways in which these vevs (and thus the symmetry breaking scale) could
be determined:\\
\indent (i) In the conventional way, radiative corrections to the scalar
potential in the presence of soft supersymmetry breaking generate a global
minimum of the potential for values of the vevs slightly below the scale
where supersymmetry breaking effects are first felt in the observable sector
\cite{faspects}. If the latter scale is the Planck scale (in a suitable
normalization) then $M_U\sim M_{Pl}/\sqrt{8\pi}\sim 10^{18}\GeV$.\\
\indent (ii) In string-derived models a pseudo $U_A(1)$ anomaly arises as a
consequence of truncating the theory to just the massless degrees of freedom,
and adds a contribution to its $D$-term, $D_A=\sum q^A_i|\vev{\phi_i}|^2
+\epsilon$, with $\epsilon=g^2{\rm Tr\,}U_A(1)/192\pi^2\sim(10^{18}\GeV)^2$
\cite{jhreview}. To avoid a huge breaking of supersymmetry we need to demand
$D_A=0$ and therefore the fields charged under $U_A(1)$ need to get suitable
vevs. Among these one generally finds the symmetry breaking Higgs fields, and
thus $M_U\sim10^{18}\GeV$ follows.\\
\indent In general, both these mechanisms could produce somewhat lower values
of $M_U$. However, $M_U\gsim10^{16}\GeV$ is necessary to avoid too rapid proton
decay due to dimension-six operators \cite{EKN}. In these more general cases
the $SU(5)$ and $U(1)$ gauge couplings would not unify at $M_U$ (only
$\alpha_2$ and $\alpha_3$ would), although they would eventually
``superunify" at the string scale $M_{SU}\sim10^{18}\GeV$. To simplify matters,
below we consider the simplest possible case of $M_U=M_{SU}\sim10^{18}\GeV$.
We also draw inspiration from string model-building and regard the Higgs
mixing term $\mu h\bar h$ as a result of an effective higher-order coupling
\cite{decisive,muproblem,Casasmu}, instead of as a result of a light singlet
field getting a small vev (\ie, $\lambda h\bar h\phi\to\lambda\vev{\phi}h\bar
h$) as originally considered \cite{revitalized,faspects}. An additional
contribution to $\mu$ is also generically present in supergravity models
\cite{GM,Casasmu,KL}.

The choice of supersymmetry breaking scenario is less clear. Below we show
that the phenomenologically acceptable choices basically fall in two
categories:
\begin{enumerate}
\item The no-scale ansatz \cite{LN}, which ensures the vanishing of the
(tree-level) cosmological constant even after supersymmetry breaking. This
framework also arises in the low-energy limit of superstring theory
\cite{Witten}. In a theory which contains heavy fields, the minimal no-scale
structure $SU(1,1)$ \cite{nsI} is generalized to $SU(N,1)$ \cite{nsII} which
implies that the scalar fields do not feel the supersymmetry breaking effects.
In practice this means that the universal scalar mass ($m_0$) and the universal
cubic scalar coupling ($A$) are set to zero. The sole source of supersymmetry
breaking is the universal gaugino mass ($m_{1/2}$), \ie,
\beq
m_0=0,\qquad A=0.\label{nsc}
\eeq
\item The dilaton $F$-term scenario, which also leads to universal soft
supersymmetry breaking parameters \cite{KL}
\beq
m_0=\coeff{1}{\sqrt{3}}m_{1/2},\qquad A=-m_{1/2}.\label{kl}
\eeq
\end{enumerate}
In either case, after enforcement of the
above constraints, the low-energy theory can be described in terms of just
three parameters: the top-quark mass ($m_t$), the ratio of Higgs vacuum
expectation values ($\tan\beta$), and the gluino mass ($m_{\tilde g}\propto
m_{1/2}$). Therefore, measurement of only two sparticle or Higgs masses would
determine the remaining thirty. Moreover, if the hidden sector responsible for
these patterns of soft supersymmetry breaking is specified, the gravitino mass
will also be determined and the supersymmetry breaking sector of the theory
will be completely fixed.

In sum, we see basically two {\em unified} string supergravity models emerging
as good candidates for  phenomenologically acceptable string models, both of
which include a flipped $SU(5)$ observable gauge group supplemented by matter
representations in order to unify at the string scale $M_U\sim10^{18}\GeV$
\cite{price,sism}, and supersymmetry breaking is parametrized by either of
the scenaria in Eqs. (\ref{nsc},\ref{kl}).

We should remark that a real string model will include a hidden sector in
addition to the observable sector discussed in what follows. The model
presented here tacitly assumes that such hidden sector is present and that
it has suitable properties. For example, the superpotential in Eq. (\ref{W})
below, in a string model will receive contributions from cubic and higher-order
terms, with the latter generating effective observable sector couplings
once hidden sector matter condensates develop \cite{decisive}. The hidden
sector is also assumed to play a fundamental role in triggering supersymmetry
breaking via \eg, gaugino condensation. This in turn would make possible the
mechanism for gauge symmetry breaking discussed above. Probably the most
important constraint on this sector of the theory is that it should yield
one of the two supersymmetry breaking scenaria outlined above.

This paper is organized as follows. In Sec. 2 we present the string-inspired
model with all the model-building details which determine in principle the
masses of the new heavy vector-like particles. We also discuss the question
of the possible re-introduction of dangerous dimension-five proton decay
operators in this generalized model. We then impose the constraint of flipped
$SU(5)$ unification and string unification to occur at $M_U=10^{18}\GeV$
to deduce the unknown masses. In Sec. 3 we discuss the various supersymmetry
breaking scenaria. In Sec. 4 we consider the experimental predictions for all
the sparticle and one-loop corrected Higgs boson masses in these models, and
deduce several simple relations among the various sparticle masses. In Sec. 5
we repeat this analysis for special more constrained cases of the chosen
supersymmetry breaking scenaria. In Sec. 6 we discuss the prospects for
experimental detection of these particles at Fermilab, LEPI,II, and HERA.
Finally, in Sec. 7 we summarize our conclusions.

\section{The Model: Gauge-Matter Structure and Properties}
The model we consider is a generalization of that presented in
Ref. \cite{revitalized}, and contains the following flipped $SU(5)$ fields:
\begin{enumerate}
\item three generations of quark and lepton fields $F_i,\bar f_i,l^c_i,\,
i=1,2,3$;
\item two pairs of Higgs \r{10},\rb{10} representations $H_i,\bar H_i,\,
i=1,2$;
\item one pair of ``electroweak" Higgs \r{5},\rb{5} representations
$h,\bar h$;
\item three singlet fields $\phi_{1,2,3}$.
\end{enumerate}
\noindent Under $SU(3)\times SU(2)$ the various flipped $SU(5)$ fields
decompose as follows:
\begin{eqnarray}
F_i&=&\{Q_i,d^c_i,\nu^c_i\},\quad \bar f_i=\{L_i,u^c_i\},
\quad l^c_i=e^c_i,\\
H_i&=&\{Q_{H_i},d^c_{H_i},\nu^c_{H_i}\},\quad
\bar H_i=\{Q_{\bar H_i},d^c_{\bar H_i},\nu^c_{\bar H_i}\},\\
h&=&\{H,D\},\quad \bar h=\{\bar H,\bar D\}.
\end{eqnarray}
The most general effective\footnote{To be understood in the string context as
arising from cubic and higher order terms \cite{KLN,decisive}.}
superpotential consistent with $SU(5)\times U(1)$ symmetry is given by
\begin{eqnarray}
W&=&\lambda^{ij}_1 F_iF_jh+\lambda^{ij}_2 F_i\bar f_j \bar h
+\lambda^{ij}_3 \bar f_il^c_j h +\mu h\bar h
+\lambda^{ij}_4 H_iH_jh+\lambda^{ij}_5\bar H_i\bar H_j\bar h\nonumber\\
&+&\lambda^{ij}_{1'}H_iF_jh+\lambda^{ij}_{2'}H_i\bar f_j\bar h
+\lambda^{ijk}_6 F_i\bar H_j\phi_k+w^{ij}H_i\bar H_j+\mu^{ij}\phi_i\phi_j.
\label{W}
\end{eqnarray}
Symmetry breaking is effected by non-zero vevs $\vev{\nu^c_{H_i}}=V_i$,
$\vev{\nu^c_{\bar H_i}}=\bar V_i$, such that
$V^2_1+V^2_2=\bar V^2_1+\bar V^2_2$.
\subsection{Higgs doublet and triplet mass matrices}
The Higgs doublet mass matrix receives contributions from
$\mu h\bar h\to \mu H\bar H$ and $\lambda^{ij}_{2'}H_i\bar f_j\bar h\to
\lambda^{ij}_{2'}V_i L_j\bar H$. The resulting matrix is
\beq
{\cal M}_2=\bordermatrix{&\bar H\cr H&\mu\cr
L_1&\lambda^{i1}_{2'}V_i\cr
L_2&\lambda^{i2}_{2'}V_i\cr
L_3&\lambda^{i3}_{2'}V_i\cr}.
\eeq
To avoid fine-tunings of the $\lambda^{ij}_{2'}$ couplings we must demand
$\lambda^{ij}_{2'}\equiv0$, so that $\bar H$ remains light.

The Higgs triplet matrix receives several contributions:
$\mu h\bar h\to\mu D\bar D$;
$\lambda^{ij}_{1'}H_iF_jh\to\lambda^{ij}_{1'}V_id^c_jD$;
$\lambda^{ij}_4 H_iH_jh\to\lambda^{ij}_4V_i d^c_{H_j}D$;
$\lambda^{ij}_5 \bar H_i\bar H_j\bar h
\to\lambda^{ij}_5\bar V_i d^c_{\bar H_j}\bar D$;
$w^{ij}d^c_{H_i}d^c_{\bar H_j}$. The resulting matrix is\footnote{The zero
entries in ${\cal M}_3$ result from the assumption $\vev{\phi_k}=0$ in
$\lambda_6^{ijk}F_i\bar H_j\phi_k$.}
\beq
{\cal M}_3=\bordermatrix{
&\bar D&d^c_{H_1}&d^c_{H_2}&d^c_1&d^c_2&d^c_3\cr
D&\mu&\lambda^{i1}_4V_i&\lambda^{i2}_4V_i&\lambda^{i1}_{1'}V_i
&\lambda^{i2}_{1'}V_i&\lambda^{i3}_{1'}V_i\cr
d^c_{\bar H_1}&\lambda^{i1}_5\bar V_i&w_{11}&w_{12}&0&0&0\cr
d^c_{\bar H_2}&\lambda^{i2}_5\bar V_i&w_{21}&w_{22}&0&0&0\cr}.\label{IIb}
\eeq
Clearly three linear combinations of $\{\bar D,d^c_{H_{1,2}},d^c_{1,2,3}\}$
will remain light. In fact, such a general situation will induce a mixing
in the down-type Yukawa matrix $\lambda^{ij}_1 F_iF_jh\to\lambda^{ij}_1Q_i
d^c_jH$, since the $d^c_j$ will need to be re-expressed in terms of these
mixed light eigenstates.\footnote{Note that this mixing is on top of any
structure that $\lambda^{ij}_1$ may have, and is the only source of mixing in
the typical string model-building case of a diagonal $\lambda_2$ matrix.} This
low-energy quark-mixing mechanism is an explicit realization of the general
extra-vector-abeyance (EVA) mechanism of Ref. \cite{EVA}. As a first
approximation though, in what follows we will set $\lambda^{ij}_{1'}=0$, so
that the light eigenstates are $d^c_{1,2,3}$.
\subsection{Neutrino see-saw matrix}
The see-saw neutrino matrix receives contributions from:
$\lambda^{ij}_2F_i\bar f_j\bar h\to m^{ij}_u\nu^c_i\nu_j$;
$\lambda^{ijk}_6 F_i\bar H_j\phi_k\to\lambda^{ijk}_6\bar V_j\nu^c_i\phi_k$;
$\mu^{ij}\phi_i\phi_j$. The resulting matrix is\footnote{We neglect a possible
higher-order contribution which could produce a non-vanishing $\nu^c_i\nu^c_j$
entry \cite{chorus}.}
\beq
{\cal M}_\nu=\bordermatrix{&\nu_j&\nu^c_j&\phi_j\cr
\nu_i&0&m^{ji}_u&0\cr
\nu^c_i&m^{ij}_u&0&\lambda^{ikj}_6\bar V_k\cr
\phi_i&0&\lambda^{jki}_6\bar V_k&\mu^{ij}\cr}.
\eeq
\subsection{Numerical scenario}
To simplify the discussion we will assume, besides\footnote{In Ref.
\cite{revitalized} the discrete symmetry $H_1\to -H_1$ was imposed so that
these couplings automatically vanish when $H_2,\bar H_2$ are not present. This
symmetry (generalized to $H_i\to -H_i$) is not needed here since it would
imply $w^{ij}\equiv0$, which is shown below to be disastrous for gauge
coupling unification.} $\lambda^{ij}_{1'}=\lambda^{ij}_{2'}\equiv0$, that
\begin{eqnarray}
\lambda^{ij}_4&=&\delta^{ij}\lambda^{(i)}_4,\quad
\lambda^{ij}_5=\delta^{ij}\lambda^{(i)}_5,\quad
\lambda^{ijk}_6=\delta^{ij}\delta^{ik}\lambda^{(i)}_6,\\
\mu^{ij}&=&\delta^{ij}\mu_i,\quad w^{ij}=\delta^{ij}w_i.
\end{eqnarray}
These choices are likely to be realized in string versions of this model
and will not alter our conclusions below. In this case the Higgs triplet mass
matrix reduces to
\beq
{\cal M}_3=\bordermatrix
{&\bar D&d^c_{H_1}&d^c_{H_2}\cr
D&\mu&\lambda^{(1)}_4V_1&\lambda^{(2)}_4V_2\cr
d^c_{\bar H_1}&\lambda^{(1)}_5\bar V_1&w_1&0\cr
d^c_{\bar H_2}&\lambda^{(2)}_5\bar V_2&0&w_2\cr}.\label{III}
\eeq
Regarding the $(3,2)$ states, the scalars get either eaten by the $X,Y$ $SU(5)$
heavy gauge bosons or become heavy Higgs bosons, whereas the fermions interact
with the $\wt X,\wt Y$ gauginos through the following mass matrix \cite{LNY}
\beq
{\cal M}_{(3,2)}=\bordermatrix
{&Q_{\bar H_1}&Q_{\bar H_2}&\wt Y\cr
Q_{H_1}&w_1&0&g_5V_1\cr
Q_{H_2}&0&w_2&g_5V_2\cr
\wt X&g_5\bar V_1&g_5\bar V_2&0\cr}.
\eeq
The lightest eigenvalues of these two matrices (denoted generally by $d^c_H$
and $Q_H$ respectively) constitute the new relatively light particles in the
spectrum, which are hereafter referred to as the ``{\em gap}" particles since
with suitable masses they bridge the gap between unification masses at
$10^{16}\GeV$ and $10^{18}\GeV$.

Guided by the phenomenological requirement on the gap particle masses, \ie,
$M_{Q_H}\gg M_{d^c_H}$ \cite{sism}, we consider the following explicit
numerical scenario
\beq
\lambda^{(2)}_4=\lambda^{(2)}_5=0,\quad
V_1,\bar V_1,V_2,\bar V_2\sim V\gg w_1\gg w_2\gg\mu,\label{V}
\eeq
which would need to be reproduced in a viable string-derived model. From Eq.
(\ref{III}) we then get $M_{d^c_{H_2}}=M_{d^c_{\bar H_2}}=w_2$, and all
other mass eigenstates $\sim V$. Furthermore, ${\cal M}_{(3,2)}$ has a
characteristic polynomial $\lambda^3-\lambda^2(w_1+w_2)-\lambda(2V^2-w_1w_2)
+(w_1+w_2)V^2=0$, which has two roots of ${\cal O}(V)$ and one root of
${\cal O}(w_1)$. The latter corresponds to $\sim(Q_{H_1}-Q_{H_2})$ and
$\sim(Q_{\bar H_1}-Q_{\bar H_2})$. In sum then, the gap particles have masses
$M_{Q_H}\sim w_1$ and $M_{d^c_H}\sim w_2$, whereas all other heavy particles
have masses $\sim V$.

The see-saw matrix reduces to
\beq
{\cal M}_\nu=\bordermatrix{&\nu_i&\nu^c_i&\phi_i\cr
\nu_i&0&m^i_u&0\cr
\nu^c_i&m^i_u&0&\lambda^{(i)}\bar V_i\cr
\phi_i&0&\lambda^{(i)}\bar V_i&\mu^i\cr},
\eeq
for each generation. The physics of this see-saw matrix has been
discussed in Ref. \cite{chorus} and more generally in Ref. \cite{ELNO},
where it was shown to lead to an interesting amount of hot dark matter
($\nu_\tau$) and an MSW-effect ($\nu_e,\nu_\mu$) compatible with all solar
neutrino data. Moreover, the out-of-equilibrium decays of the $\nu^c$
``flipped neutrino" fields in the early Universe induce a lepton number
asymmetry which is later processed into a baryon number asymmetry by
non-perturbative electroweak processes \cite{ENO,ELNO}. All these phenomena can
occur in the same region of parameter space.
\subsection{Proton decay}
The dimension-six operators mediating proton decay in this model are highly
suppressed due to the large mass of the $X,Y$ gauge bosons
($\sim M_U=10^{18}\GeV$). Higgsino mediated dimension-five operators exist
and are naturally suppressed in the minimal model of Ref. \cite{revitalized}.
The reason for this is that the Higgs triplet mixing term
$\mu h\bar h\to \mu D\bar D$ is small ($\mu\sim M_Z$), whereas the Higgs
triplet mass eigenstates obtained from Eq. (\ref{IIb}) by just keeping the
$2\times2$ submatrix in the upper left-hand corner, are always very heavy
($\sim V$). The dimension-five mediated operators are then proportional to
$\mu/V^2$ and thus the rate is suppressed by a factor or $(\mu/V)^2\ll1$
relative to the unsuppressed case found in the standard $SU(5)$ model.

In the generalized model presented here, the Higgs triplet mixing term is
still $\mu D\bar D$. However, the exchanged mass eigenstates are not
necessarily all very heavy. In fact, above we have demanded the existence of a
relatively light ($\sim w_1$) Higgs triplet state ($d^c_H$). In this case
the operators are proportional to $\mu\alpha_i\bar\alpha_i/{\cal M}^2_i$, where
${\cal M}_i$ is the mass of the $i$-th exchanged eigenstate and
$\alpha_i,\bar\alpha_i$ are its $D,\bar D$ admixtures. In the scenario
described above, the relatively light eigenstates ($d^c_{H_2},d^c_{\bar H_2}$)
contain no $D,\bar D$ admixtures, and the operator will again be
$\propto\mu/V^2$.

Note however that if conditions (\ref{V}) (or some analogous suitability
requirement) are not satisfied, then diagonalization of ${\cal M}_3$ in Eq.
(\ref{III}) may re-introduce a sizeable dimension-five mediated proton decay
rate, depending on the value of the $\alpha_i,\bar\alpha_i$ coefficients. To be
safe one should demand \cite{dfive,LNZ}
\beq
{\mu\alpha_i\bar\alpha_i\over {\cal M}^2_i}\lsim{1\over 10^{17}\GeV}.
\eeq
For the higher values of $M_{d^c_H}$ in Table \ref{Table1} (see below), this
constraint can be satisfied for not necessarily small values of
$\alpha_i,\bar\alpha_i$.
\subsection{Gauge coupling unification}
Since we have chosen $V\sim M_U=M_{SU}=10^{18}\GeV$, this means that the
Standard Model gauge couplings should unify at the scale $M_U$. However, their
running will be modified due to the presence of the gap particles. Note that
the underlying flipped $SU(5)$ symmetry, even though not evident in this
respect, is nevertheless essential in the above discussion. The masses $M_Q$
and $M_{d^c_H}$ can then be determined, as follows \cite{sism}
\begin{eqnarray}
\ln{M_{Q_H}\over m_Z}&=&\pi\left({1\over2\alpha_e}-{1\over3\alpha_3}
-{\sin^2\theta_w-0.0029\over\alpha_e}\right)-2\ln{M_U\over m_Z}-0.63,\\
\ln{M_{d^c_H}\over m_Z}&=&\pi\left({1\over2\alpha_e}-{7\over3\alpha_3}
+{\sin^2\theta_w-0.0029\over\alpha_e}\right)
			-6\ln{M_U\over m_Z}-1.47,\label{VIb}
\end{eqnarray}
where $\alpha_e$, $\alpha_3$ and $\sin^2\theta_w$ are all measured at $M_Z$.
This is a one-loop determination (the constants account for the dominant
two-loop corrections) which neglects all low- and high-energy threshold
effects,\footnote{Here we assume a common supersymmetric threshold at $M_Z$.
In fact, the supersymmetric threshold and the $d^c_H$ mass are anticorrelated.
See Ref. \cite{sism} for a discussion.} but is quite adequate for our present
purposes. As shown in Table \ref{Table1} (and Eq. (\ref{VIb})) the $d^c_H$ mass
depends most sensitively on $\alpha_3(M_Z)=0.118\pm0.008$ \cite{Bethke},
whereas the $Q_H$ mass and the unified coupling are rather insensitive to it.
The unification of the gauge couplings is shown in Fig. \ref{Figure1} (solid
lines) for the central value of $\alpha_3(M_Z)$. This figure also shows the
case of no gap particles (dotted lines), for which $M_U\approx10^{16}\GeV$.

\begin{table}
\hrule
\caption{The value of the gap particle masses and the unified coupling for
$\alpha_3(M_Z)=0.118\pm0.008$. We have taken $M_U=10^{18}\GeV$,
$\sin^2\theta_w=0.233$, and $\alpha^{-1}_e=127.9$.}
\label{Table1}
\begin{center}
\begin{tabular}{|c|c|c|c|}\hline
$\alpha_3(M_Z)$&$M_{d^c_H}\,(\GeV)$&$M_{Q_H}\,(\GeV)$&$\alpha(M_U)$\\ \hline
$0.110$&$4.9\times10^4\GeV$&$2.2\times10^{12}\GeV$&$0.0565$\\
$0.118$&$4.5\times10^6\GeV$&$4.1\times10^{12}\GeV$&$0.0555$\\
$0.126$&$2.3\times10^8\GeV$&$7.3\times10^{12}\GeV$&$0.0547$\\ \hline
\end{tabular}
\end{center}
\hrule
\end{table}

\begin{figure}
\vspace{4in}
\vspace{-0.7in}
\caption{\baselineskip=12pt
The running of the gauge couplings in the flipped $SU(5)$ model for
$\alpha_3(M_Z)=0.118$ (solid lines). The gap particle masses have been derived
using the gauge coupling RGEs to achieve unification at $M_U=10^{18}\GeV$. The
case with no gap particles (dotted lines) is also shown; here
$M_U\approx10^{16}\GeV$.}
\label{Figure1}
\end{figure}

\section{The Model: Supersymmetry Breaking Scenaria}
Supersymmetry breaking in string models can generally be triggered in a
phenomenologically acceptable way by non-zero $F$-terms for: (a) any of the
moduli fields of the string model ($\vev{F_M}$) \cite{FM}, (b) the dilaton
field ($\vev{F_D}$) \cite{KL}, or (c) the hidden matter fields ($\vev{F_H}$)
\cite{AELN}. It has been recently noted  \cite{KL} that much model-independent
information can be obtained about the structure of the soft supersymmetry
breaking parameters in generic string supergravity models if one neglects the
third possibility ($\vev{F_H}=0$) and assumes that either: (i)
$\vev{F_M}\gg\vev{F_D}$, or (ii) $\vev{F_D}\gg\vev{F_M}$.

In case (i) the scalar masses are generally not universal, \ie, $m_i=f_im_0$
where $m_0$ is the gravitino mass and $f_i$ are calculable constants,
and therefore large flavor-changing-neutral-currents (FCNCs) \cite{EN} are
potentially dangerous \cite{IL}. The gaugino masses arise from the one-loop
contribution to the gauge kinetic function and are thus suppressed
($m_{1/2}\sim(\alpha/4\pi)m_0$) \cite{IL,Casas,KL}. The experimental
constraints on the gaugino masses then force the squark and slepton masses
into the TeV range \cite{Casas}. It is interesting to note that this
supersymmetry breaking scenario is not unlike that required for the minimal
$SU(5)$ supergravity model in order to have the dimension-five proton decay
operators under control \cite{dfive,LNZ}, which requires
$m_{1/2}/m_0\lsim\coeff{1}{3}$. This constraint entails potential cosmological
troubles: the neutralino relic density is large and one needs to tune the
parameters to have the neutralino mass be very near the Higgs and $Z$
resonances \cite{LNZ,ANcosm,poles}. Clearly, such cosmological constraints are
going to be exacerbated in the case (i) scenario ($m_{1/2}/m_0\ll1$) and will
likely require real fine-tuning of the model parameters.

An important exception to case (i) occurs if $f_i\equiv0$ and all scalar masses
at the unification scale vanish ($\vev{F_M}_{m_0=0}$), as is the case in
unified no-scale supergravity models \cite{LN}. This special case automatically
restores the much needed universality of scalar masses, and in the context of
no-scale models also entails $A=0$, see Eq. (\ref{nsc}). A special case of this
scenario occurs when the bilinear soft-supersymmetry breaking mass parameter
$B(M_U)$ is also required to vanish. With the additional ingredient of a
flipped $SU(5)$ gauge group, all the above problems are naturally avoided
\cite{LNZb}, and interesting predictions for direct
\cite{LNWZ,LNPWZh,LNPWZ,hera} and indirect \cite{bsgamma,ewcorr,NT}
experimental detection follow.

If supersymmetry breaking is triggered by $\vev{F_D}$ (case (ii)), one obtains
{\em universal} soft-supersymmetry gaugino and scalar masses and trilinear
interactions \cite{KL} and the soft-supersymmetry breaking parameters
in Eq. (\ref{kl}) result. As well, there is a special more constrained case
where $B(M_U)=2m_0={2\over\sqrt{3}}m_{1/2}$ is also required, if one demands
that the $\mu$ parameter receive contributions solely from supergravity
\cite{KL}. With the complement of a flipped $SU(5)$ structure, this model has
also been seen to avoid all the difficulties of the generic $\vev{F_M}$
scenario \cite{lnzII}. This supersymmetry breaking scenario has been studied
recently also in the context of the minimal supersymmetric Standard Model
(MSSM) in Ref. \cite{BLM}.

Therefore, in what follows we restrict ourselves to the two supersymmetry
breaking scenaria in Eqs. (\ref{nsc},\ref{kl}) and their special cases
($B(M_U)=0$ and $B(M_U)=2m_0$, respectively).

\section{Phenomenology: General Case}
The procedure to extract the low-energy predictions of the models outlined
above is rather standard by now (see \eg, Ref. \cite{aspects}): (a) the
bottom-quark and tau-lepton masses, together with the input values of $m_t$ and
$\tan\beta$ are used to determine the respective Yukawa couplings at the
electroweak scale; (b) the gauge and Yukawa couplings are then run up to the
unification scale $M_U=10^{18}\GeV$ taking into account the extra vector-like
quark doublet ($\sim10^{12}\GeV$) and singlet ($\sim10^6\GeV$) introduced above
\cite{sism,LNZb}; (c) at the unification scale the soft-supersymmetry breaking
parameters are introduced (according to Eqs. (\ref{nsc},\ref{kl})) and the
scalar masses are then run down to the electroweak scale; (d) radiative
electroweak symmetry breaking is enforced by minimizing the one-loop effective
potential which depends on the whole mass spectrum, and the values of the Higgs
mixing term $|\mu|$ and the bilinear soft-supersymmetry breaking parameter $B$
are determined from the minimization conditions; (e) all known phenomenological
constraints on the sparticle and Higgs masses are applied (most importantly the
LEP lower bounds on the chargino and Higgs masses), including the cosmological
requirement of not-too-large neutralino relic density.
\subsection{Mass ranges}
We have scanned the parameter space for $m_t=130,150,170\GeV$,
$\tan\beta=2\to50$ and $m_{1/2}=50\to500\GeV$. Imposing the constraint
$m_{\tilde g,\tilde q}<1\TeV$ we find
\begin{eqnarray}
&\vev{F_M}_{m_0=0}:\qquad	&m_{1/2}<475\GeV,\quad \tan\beta\lsim32,\\
&\vev{F_D}:\qquad	&m_{1/2}<465\GeV,\quad \tan\beta\lsim46.
\end{eqnarray}
These restrictions on $m_{1/2}$ cut off the growth of most of the sparticle and
Higgs masses at $\approx1\TeV$. However, the sleptons, the lightest Higgs, the
two lightest neutralinos, and the lightest chargino are cut off at a much lower
mass, as follows\footnote{In this class of supergravity models the three
sneutrinos ($\tilde\nu$) are degenerate in mass. Also, $m_{\tilde
\mu_L}=m_{\tilde e_L}$ and $m_{\tilde\mu_R}=m_{\tilde e_R}$.}
\begin{eqnarray}
&\vev{F_M}_{m_0=0}:&\left\{
	\begin{array}{l}
	m_{\tilde e_R}<190\GeV,\quad m_{\tilde e_L}<305\GeV,
				\quad m_{\tilde\nu}<295\GeV\\
	m_{\tilde\tau_1}<185\GeV,\quad m_{\tilde\tau_2}<315\GeV\\
	m_h<125\GeV\\
	m_{\chi^0_1}<145\GeV,\quad m_{\chi^0_2}<290\GeV,
				\quad m_{\chi^\pm_1}<290\GeV
	\end{array}
		\right.\\
&\vev{F_D}:&\left\{
	\begin{array}{l}
	m_{\tilde e_R}<325\GeV,\quad m_{\tilde e_L}<400\GeV,
				\quad m_{\tilde\nu}<400\GeV\\
	m_{\tilde\tau_1}<325\GeV,\quad m_{\tilde\tau_2}<400\GeV\\
	m_h<125\GeV\\
	m_{\chi^0_1}<145\GeV,\quad m_{\chi^0_2}<285\GeV,
				\quad m_{\chi^\pm_1}<285\GeV
	\end{array}
		\right.
\end{eqnarray}
It is interesting to note that due to the various constraints on the model,
the gluino and (average) squark masses are bounded from below,
\beq
\vev{F_M}_{m_0=0}:\left\{
	\begin{array}{l}
	m_{\tilde g}\gsim245\,(260)\GeV\\
	m_{\tilde q}\gsim240\,(250)\GeV
	\end{array}
		\right.
\qquad
\vev{F_D}:\left\{
	\begin{array}{l}
	m_{\tilde g}\gsim195\,(235)\GeV\\
	m_{\tilde q}\gsim195\,(235)\GeV
	\end{array}
		\right.		\label{gmin}
\eeq
for $\mu>0(\mu<0)$. Relaxing the above conditions on $m_{1/2}$ simply allows
all sparticle masses to grow further proportional to $m_{\tilde g}$.

\begin{table}
\hrule
\caption{
The value of the $c_i$ coefficients appearing in  Eq.~(25), the ratio
$c_{\tilde g}=m_{\tilde g}/m_{1/2}$, and the average squark coefficient
$\bar c_{\tilde q}$, for $\alpha_3(M_Z)=0.118\pm0.008$. Also shown are the
$a_i,b_i$ coefficients for the central value of $\alpha_3(M_Z)$ and both
supersymmetry breaking scenaria ($M:\vev{F_M}_{m_0=0}$, $D:\vev{F_D}$).
 The results apply as well to the second-generation squark and slepton masses.}
\label{Table2}
\begin{center}
\begin{tabular}{|c|c|c|c|}\hline
$i$&$c_i\,(0.110)$&$c_i\,(0.118)$&$c_i\,(0.126)$\\ \hline
$\tilde\nu,\tilde e_L$&$0.406$&$0.409$&$0.413$\\
$\tilde e_R$&$0.153$&$0.153$&$0.153$\\
$\tilde u_L,\tilde d_L$&$3.98$&$4.41$&$4.97$\\
$\tilde u_R$&$3.68$&$4.11$&$4.66$\\
$\tilde d_R$&$3.63$&$4.06$&$4.61$\\
$c_{\tilde g}$&$1.95$&$2.12$&$2.30$\\
$\bar c_{\tilde q}$&$3.82$&$4.07$&$4.80$\\ \hline
\end{tabular}
\begin{tabular}{|c|c|c|c|c|}\hline
$i$&$a_i(M)$&$b_i(M)$&$a_i(D)$&$b_i(D)$\\ \hline
$\tilde e_L$&$0.302$&$+1.115$&$0.406$&$+0.616$\\
$\tilde e_R$&$0.185$&$+2.602$&$0.329$&$+0.818$\\
$\tilde\nu$&$0.302$&$-2.089$&$0.406$&$-1.153$\\
$\tilde u_L$&$0.991$&$-0.118$&$1.027$&$-0.110$\\
$\tilde u_R$&$0.956$&$-0.016$&$0.994$&$-0.015$\\
$\tilde d_L$&$0.991$&$+0.164$&$1.027$&$+0.152$\\
$\tilde d_R$&$0.950$&$-0.033$&$0.989$&$-0.030$\\ \hline
\end{tabular}
\end{center}
\hrule
\end{table}

\subsection{Mass relations}
The neutralino and chargino masses show a correlation observed before in
this class of models \cite{ANc,LNZb}, namely
\beq
m_{\chi^0_1}\approx \coeff{1}{2}m_{\chi^0_2},\qquad
m_{\chi^0_2}\approx m_{\chi^\pm_1}\approx M_2=(\alpha_2/\alpha_3)m_{\tilde g}
\approx0.28m_{\tilde g}.\label{neuchar}
\eeq
This is because throughout the parameter space $|\mu|$ is generally much larger
than $M_W$ and $|\mu|>M_2$. In practice we find $m_{\chi^0_2}\approx
m_{\chi^\pm_1}$ to be satisfied quite accurately, whereas
$m_{\chi^0_1}\approx{1\over2}m_{\chi^0_2}$ is only qualitatively satisfied,
although the agreement is better in the $\vev{F_D}$ case. In fact, these two
mass relations are much more reliable than the one that links them to
$m_{\tilde g}$. The heavier neutralino ($\chi^0_{3,4}$) and chargino
($\chi^\pm_2$) masses are determined by the value of $|\mu|$; they all approach
this limit for large enough $|\mu|$. More precisely, $m_{\chi^0_3}$ approaches
$|\mu|$ sooner than $m_{\chi^0_4}$ does. On the other hand, $m_{\chi^0_4}$
approaches $m_{\chi^\pm_2}$ rather quickly.

The first- and second-generation squark and slepton masses can be determined
analytically
\beq
\wt m_i=\left[m^2_{1/2}(c_i+\xi^2_0)-d_i{\tan^2\beta-1\over\tan^2\beta+1}
M^2_W\right]^{1/2}=a_i m_{\tilde g}\left[1+b_i\left({150\over m_{\tilde
g}}\right)^2{\tan^2\beta-1\over\tan^2\beta+1}\right]^{1/2},\label{masses}
\eeq
where $d_i=(T_{3i}-Q)\tan^2\theta_w+T_{3i}$ (\eg, $d_{\tilde u_L}={1\over2}
-{1\over6}\tan^2\theta_w$, $d_{\tilde e_R}=-\tan^2\theta_w$), and
$\xi_0=m_0/m_{1/2}=0,\coeff{1}{\sqrt{3}}$. The coefficients $c_i$ can be
calculated numerically in terms of the low-energy gauge couplings, and are
given in  Table \ref{Table2}\footnote{These are renormalized at the scale
$M_Z$. In a more accurate treatment, the $c_i$ would be renormalized at the
physical sparticle mass scale, leading to second order shifts on the sparticle
masses.} for $\alpha_3(M_Z)=0.118\pm0.008$. In the table we also give
$c_{\tilde g}=m_{\tilde g}/m_{1/2}$. Note that these values are smaller than
what is obtained in the minimal $SU(5)$ supergravity model (where $c_{\tilde
g}=2.90$ for $\alpha_3(M_Z)=0.118$) and therefore the numerical relations
between the gluino mass and the neutralino masses are different in that model.
In the table we also show the resulting values for $a_i,b_i$ for the central
value of $\alpha_3(M_Z)$. Note that the apparently larger $\tan\beta$
dependence in the $\vev{F_M}_{m_0=0}$ case (\ie, $|b_i(M)|>|b_i(D)|$) is
actually compensated by a larger minimum value of $m_{\tilde g}$ in this case
(see Eq. (\ref{gmin})).

The ``average" squark mass, $m_{\tilde q}\equiv{1\over8}(m_{\tilde
u_L}+m_{\tilde u_R}+m_{\tilde d_L}+m_{\tilde d_R}+m_{\tilde c_L}+m_{\tilde c_R}
+m_{\tilde s_L}+m_{\tilde s_R})
=(m_{\tilde g}/c_{\tilde q})\sqrt{\bar c_{\tilde q}+\xi^2_0}$, with $\bar
c_{\tilde q}$ given in Table \ref{Table2}, is determined to be
\beq
m_{\tilde q}=\left\{	\begin{array}{ll}
			(1.00,0.95,0.95) m_{\tilde g},&\quad\vev{F_M}_{m_0=0}\\
			(1.05,0.99,0.98) m_{\tilde g},&\quad\vev{F_D}
			\end{array}
		\right.
\eeq
for $\alpha_3(M_Z)=0.110,0.118,0.126$ (the dependence on $\tan\beta$ is small).
The squark splitting around the average is $\approx2\%$.

These masses are plotted in Fig. \ref{Figure2}. The thickness and straightness
of the lines shows the small $\tan\beta$ dependence, except for $\tilde\nu$.
The results do not depend on the sign of $\mu$, except to the extent that some
points in parameter space are not allowed for both signs of $\mu$: the $\mu<0$
lines start-off at larger mass values. Note that
\beq
\vev{F_M}_{m_0=0}:\left\{
	\begin{array}{l}
	m_{\tilde e_R}\approx0.18m_{\tilde g}\\
	m_{\tilde e_L}\approx0.30m_{\tilde g}\\
	m_{\tilde e_R}/m_{\tilde e_L}\approx0.61
	\end{array}
		\right.
\qquad
\vev{F_D}:\left\{
	\begin{array}{l}
	m_{\tilde e_R}\approx0.33m_{\tilde g}\\
	m_{\tilde e_L}\approx0.41m_{\tilde g}\\
	m_{\tilde e_R}/m_{\tilde e_L}\approx0.81
	\end{array}
		\right.
\eeq

\begin{figure}[p]
\vspace{4.7in}
\vspace{3.8in}
\vspace{-0.7in}
\caption{\baselineskip=12pt
The first-generation squark and slepton masses as a function of
the gluino mass, for both signs of $\mu$, $m_t=150\GeV$, and both supersymmetry
breaking scenaria under consideration. The same values apply to the second
generation. The thickness of the lines and their deviation from linearity are
because of the small $\tan\beta$ dependence.}
\label{Figure2}
\end{figure}

The third generation squark and slepton masses cannot be determined
analytically. In Fig. \ref{Figure3} we show
$\tilde\tau_{1,2},\tilde b_{1,2},\tilde t_{1,2}$ for the choice $m_t=150\GeV$.
The variability on the $\tilde\tau_{1,2}$ and $\tilde b_{1,2}$ masses
is due to the $\tan\beta$-dependence in the off-diagonal element of the
corresponding $2\times2$ mass matrices ($\propto
m_{\tau,b}(A_{\tau,b}+\mu\tan\beta)$). The off-diagonal element in the
stop-squark mass matrix ($\propto m_t(A_t+\mu/\tan\beta)$) is
rather insensitive to $\tan\beta$ but still effects a large $\tilde t_1-\tilde
t_2$ mass splitting because of the significant $A_t$ contribution. Note
that both these effects are more pronounced for the $\vev{F_D}$ case since
there $|A_{t,b,\tau}|$ are larger than in the $\vev{F_M}_{m_0=0}$ case.
The lowest values of the $\tilde t_1$ mass go up with $m_t$ and can be as low
as
\beq
m_{\tilde t_1}\gsim\left\{	\begin{array}{ll}
		160,170,190\,(155,150,170)\GeV;&\quad\vev{F_M}_{m_0=0}\\
			88,112,150\,(92,106,150)\GeV;&\quad\vev{F_D}
			\end{array}
		\right.
\eeq
for $m_t=130,150,170\GeV$ and $\mu>0\,(\mu<0)$.

\begin{figure}[p]
\vspace{4.3in}
\vspace{3.8in}
\vspace{-0.5in}
\caption{\baselineskip=12pt
The $\tilde\tau_{1,2}$, $\tilde b_{1,2}$, and $\tilde t_{1,2}$ masses
versus the gluino mass for both signs of $\mu$, $m_t=150\GeV$, and both
supersymmetry breaking scenaria. The variability in the $\tilde\tau_{1,2}$,
$\tilde b_{1,2}$, and $\tilde t_{1,2}$ masses is because of the off-diagonal
elements of the corresponding mass matrices.}
\label{Figure3}
\end{figure}

The one-loop corrected lightest CP-even ($h$) and CP-odd ($A$) Higgs boson
masses are shown in Fig. \ref{Figure4} as functions of $m_{\tilde g}$ for
$m_t=150\GeV$. Following the methods of Ref. \cite{LNPWZh} we have determined
that the LEP lower bound on $m_h$ becomes $m_h\gsim60\GeV$, as the figure
shows. The largest value of $m_h$ depends on $m_t$; we find
\beq
m_h<\left\{	\begin{array}{ll}
		106,115,125\GeV;&\quad\vev{F_M}_{m_0=0}\\
		107,117,125\GeV;&\quad\vev{F_D}
			\end{array}
		\right.
\eeq
for $m_t=130,150,170\GeV$. It is interesting to note that the one-loop
corrected values of $m_h$ for $\tan\beta=2$ are quite dependent on the sign of
$\mu$. This phenomenon can be traced back to the $\tilde t_1-\tilde t_2$ mass
splitting which enhances the dominant $\tilde t$ one-loop corrections to $m_h$
\cite{ERZ}, an effect which is usually neglected in phenomenological analyses.
The $\tilde t_{1,2}$ masses for $\tan\beta=2$ and are drawn closer together
than the rest. The opposite effect occurs for $\mu<0$ and therefore the
one-loop correction is larger in this case. The sign-of-$\mu$ dependence
appears in the off-diagonal entries in the $\tilde t$ mass matrix  $\propto
m_t(A_t+\mu/\tan\beta)$, with $A_t<0$ in this case. Clearly only small
$\tan\beta$ matters, and $\mu<0$ enhances the splitting. The $A$-mass grows
fairly linearly with $m_{\tilde g}$ with a $\tan\beta$-dependent slope which
decreases for increasing $\tan\beta$, as shown in Fig. \ref{Figure4}. Note that
even though $m_A$ can be fairly light, we always get $m_A>m_h$, in agreement
with a general theorem to this effect in supergravity theories \cite{DNh}. This
result also implies that the channel $e^+e^-\to hA$ at LEPI is not
kinematically allowed in this model.

\begin{figure}[p]
\vspace{4.3in}
\vspace{3.8in}
\vspace{-0.3in}
\caption{\baselineskip=12pt
The one-loop corrected $h$ and $A$ Higgs masses versus the gluino
mass for both signs of $\mu$, $m_t=150\GeV$, and the two supersymmetry
breaking scenaria. Representative values of $\tan\beta$ are indicated.}
\label{Figure4}
\end{figure}

\subsection{Neutralino relic density}
The computation of the neutralino relic density (following the methods of
Refs. \cite{LNYdmI,KLNPYdm}) shows that $\Omega_\chi h^2_0\lsim0.25\,(0.90)$ in
the no-scale (dilaton) model. This implies that in these models the
cosmologically interesting values $\Omega_\chi h^2_0\lsim1$ occur quite
naturally. These results are in good agreement with the observational upper
bound on $\Omega_\chi h^2_0$ \cite{KT}. Moreover, fits to the COBE data and the
small and large scale structure of the Universe suggest \cite{many} a mixture
of $\approx70\%$ cold dark matter and $\approx30\%$ hot dark matter together
with $h_0\approx0.5$. The hot dark matter component in the form of massive tau
neutrinos has already been shown to be compatible with the flipped $SU(5)$
model we consider here \cite{chorus,ELNO}, whereas the cold dark matter
component implies $\Omega_\chi h^2_0\approx0.17$ which is reachable in these
models.

\section{Phenomenology: Special Cases}
\subsection{The strict no-scale case}
We now impose the additional constraint $B(M_U)=0$ to be added to
Eq.~(\ref{nsc}), and obtain the so-called strict no-scale case. Since $B(M_Z)$
is determined by the radiative electroweak symmetry breaking conditions, this
added constraint needs to be imposed in a rather indirect way. That is, for
given $m_{\tilde g}$ and $m_t$ values, we scan the possible values of
$\tan\beta$ looking for cases where $B(M_U)=0$. The most striking result is
that solutions exist {\em only} for $m_t\lsim135\GeV$ if $\mu>0$ and for
$m_t\gsim140\GeV$ if $\mu<0$. That is, the value of $m_t$ {\em determines} the
sign of $\mu$. Furthermore, for $\mu<0$ the value of $\tan\beta$ is determined
uniquely as a function of $m_t$ and $m_{\tilde g}$, whereas for $\mu>0$,
$\tan\beta$ can be double-valued for some $m_t$ range which includes
$m_t=130\GeV$ (but does not include $m_t=100\GeV$). In Fig. \ref{Figure5} (top
row) we plot the solutions found in this manner for the indicated $m_t$
values.\footnote{The values of $\tan\beta$ shown in Fig.~\ref{Figure5} (top
row) differ somewhat from those shown previously in Ref. \cite{LNZb} (Fig. 8)
because in that paper $m_b=4.5\GeV$ was used, whereas throughout the present
paper $m_b=4.9\GeV$ is used instead.}

All the mass relationships deduced in the previous section apply here as well.
The $\tan\beta$-spread that some of them have will be much reduced though.
The most noticeable changes occur for the quantities which depend most
sensitively on $\tan\beta$. In Fig. \ref{Figure5} (bottom row) we plot the
one-loop corrected lightest Higgs boson mass versus $m_{\tilde g}$. The result
is that $m_h$ is basically determined by $m_t$; only a weak dependence on
$m_{\tilde g}$ exists. Moreover, for $m_t\lsim135\GeV\Leftrightarrow\mu>0$,
$m_h\lsim105\GeV$; whereas for $m_t\gsim140\GeV\Leftrightarrow\mu<0$,
$m_h\gsim100\GeV$. Therefore, in the strict no-scale case, once the top-quark
mass is measured, we will know the sign of $\mu$ and whether $m_h$ is above or
below $100\GeV$.

For $\mu>0$, we just showed that the strict no-scale constraint requires
$m_t\lsim135\GeV$. This implies that $\mu$ cannot grow as large as it did
previously in the general case. In fact, for $\mu>0$, $\mu_{max}\approx745\GeV$
before and $\mu_{max}\approx440\GeV$ now. This smaller value of $\mu_{max}$ has
the effect of cutting off the growth of the $\chi^0_{3,4},\chi^\pm_2$ masses
at $\approx\mu_{max}\approx440\GeV$ (c.f. $\approx750\GeV$) and of the heavy
Higgs masses at $\approx530\GeV$ (c.f. $\approx940\GeV$).

\begin{figure}[t]
\vspace{4.3in}
\vspace{-0.3in}
\caption{\baselineskip=12pt
The value of $\tan\beta$ versus $m_{\tilde g}$ in the strict no-scale case
(where $B(M_U)=0$) for the indicated values of $m_t$. Note that the sign of
$\mu$ is {\em determined} by $m_t$ and that $\tan\beta$ can be double-valued
for $\mu>0$. Also shown is the one-loop corrected lightest
Higgs boson mass. Note that if $\mu>0$ (for $m_t<135\GeV$) then $m_h<105\GeV$;
whereas if $\mu<0$ (for $m_t>140\GeV$) then $m_h>100\GeV$.}
\label{Figure5}
\end{figure}

\subsection{The special dilaton scenario case}
In our analysis above, the radiative electroweak breaking conditions were used
to determine the magnitude of the Higgs mixing term $\mu$ at the electroweak
scale. This quantity is ensured to remain light as long as the supersymmetry
breaking parameters remain light. In a fundamental theory this parameter should
be calculable and its value used to determine the $Z$-boson mass. From this
point of view it is not clear that the natural value of $\mu$ should be light.
In specific models on can obtain such values by invoking non-renormalizable
interactions \cite{muproblem,Casasmu}. Another contribution to this quantity
is generically present in string supergravity models \cite{GM,Casasmu,KL}.
The general case with contributions from both sources has been effectively
dealt with in the previous section. If one assumes that only
supergravity-induced contributions to $\mu$ exist, then it can be shown that
the $B$-parameter at the unification scale is also determined \cite{KL},
\beq
B(M_U)=2m_0=\coeff{2}{\sqrt{3}}m_{1/2},\label{klII}
\eeq
which is to be added to the set of relations in Eq. (\ref{kl}). This new
constraint effectively determines $\tan\beta$ for given $m_t$ and $m_{\tilde
g}$ values and makes this restricted version of the model highly predictive.

{}From the outset we note that only solutions with $\mu<0$ exist. This is not
a completely obvious result, but it can be partially understood as follows.
In tree-level approximation, $m^2_A>0\Rightarrow\mu B<0$ at the electroweak
scale. Since $B(M_U)$ is required to be positive and not small, $B(M_Z)$ will
likely be positive also, thus forcing $\mu$ to be negative. A sufficiently
small value of $B(M_U)$ and/or one-loop corrections to $m^2_A$ could alter this
result, although in practice this does not happen. A numerical iterative
procedure allows us to determine the value of $\tan\beta$ which satisfies Eq.
(\ref{klII}), from the calculated value of $B(M_Z)$. We find that
\beq
\tan\beta\approx1.57-1.63,1.37-1.45,1.38-1.40\quad{\rm for\ }
m_t=130,150,155\GeV
\eeq
is required. Since $\tan\beta$ is so small ($m^{tree}_h\approx28-41\GeV$), a
significant one-loop correction to $m_h$ is required to increase it above
its experimental lower bound of $\approx60\GeV$ \cite{LNPWZh}. This requires
the largest possible top-quark masses and a not-too-small squark mass. However,
perturbative unification imposes an upper bound on $m_t$ for a given
$\tan\beta$ \cite{DL}, which in this case implies \cite{aspects}
\beq
m_t\lsim155\GeV,
\eeq
which limits the magnitude of $m_h$
\beq
m_h\lsim74,87,91\GeV\qquad{\rm for}\qquad m_t=130,150,155\GeV.
\eeq
Lower values of $m_t$ are disfavored experimentally.

In Table~\ref{Table3} we give the range of sparticle and Higgs masses that
are allowed in this case. Clearly, continuing top-quark searches at the
Tevatron and Higgs searches at LEPI,II should probe this restricted scenario
completely.

\begin{table}
\hrule
\caption{
The range of allowed sparticle and Higgs masses in the restricted dilaton
scenario. The top-quark mass is restricted to be $m_t<155\GeV$. All masses in
GeV.}
\label{Table3}
\begin{center}
\begin{tabular}{|c|c|c|c|}\hline
$m_t$&$130$&$150$&$155$\\ \hline
$\tilde g$&$335-1000$&$260-1000$&$640-1000$\\
$\chi^0_1$&$38-140$&$24-140$&$90-140$\\
$\chi^0_2,\chi^\pm_1$&$75-270$&$50-270$&$170-270$\\
$\tan\beta$&$1.57-1.63$&$1.37-1.45$&$1.38-1.40$\\
$h$&$61-74$&$64-87$&$84-91$\\
$\tilde l$&$110-400$&$90-400$&$210-400$\\
$\tilde q$&$335-1000$&$260-1000$&$640-1000$\\
$A,H,H^+$&$>400$&$>400$&$>970$\\ \hline
\end{tabular}
\end{center}
\hrule
\end{table}

\section{Prospects for Experimental Detection}
The sparticle and Higgs spectrum shown in
Figs.~\ref{Figure2},\ref{Figure3},\ref{Figure4},\ref{Figure5} and
Table~\ref{Table3} can be explored partially at present and near future
collider facilities, as we discuss below for each supersymmetry breaking
scenario considered above. First, we want to point out that there are
two {\em indirect} experimental constraints which restrict these models in
a more general way \cite{bsgamma,ewcorr}: (i) the recently experimentally
determined range for the $b\to s\gamma$ rare decay mode \cite{Thorndike}
\beq
{\rm BR}(b\to s\gamma)=(0.6-5.5)\times10^{-4}
\eeq
at 95\% CL; and (ii) the precise LEP electroweak measurements which constrain
the $\epsilon_{1,2,3}$ parameters \cite{epsilons}. The first contraint is
particularly effective in removing acceptable points in parameter space since
in these models very small values of ${\rm BR}(b\to s\gamma)$ are not uncommon
\cite{bsgamma,bsgammaII}. The second constraint basically imposes an upper
bound on the top-quark mass of $\approx175\GeV$. However, for
$150\GeV<m_t<175\GeV$ a progressively stricter upper bound on the chargino mass
(and therefore on all sparticle and Higgs masses) is required, \ie,
$50\GeV<m_{\chi^\pm_1}<100\GeV$, in order to keep $\epsilon_1$ below its
current 90\% CL upper limit. This implies that the choice $m_t=170\GeV$ above
is rather constrained \cite{bsgammaII}. Setting aside these indirect
constraints on the parameter space of these models, we now discuss the
prospects for {\em direct} experimental detection.

\subsection{Tevatron}
\begin{description}
\item (a) The search and eventual discovery of the top quark will narrow down
the three-dimensional parameter space of these models considerably. Moreover,
in the two special cases discussed in the previous section this measurement
will be very important: (i) in the strict no-scale case it will determine the
sign of $\mu$ ($\mu>0$ if $m_t\lsim135\GeV$; $\mu<0$ if $m_t\gsim140\GeV$)
and whether the Higgs mass is above or below $\approx100\GeV$, and (ii) it may
rule out the restricted dilaton scenario if $m_t>150\GeV$.
\item (b) The trilepton signal in $p\bar p\to \chi^0_2\chi^\pm_1X$, where
$\chi^0_2$ and $\chi^\pm_1$ both decay leptonically, is a clean test of
supersymmetry \cite{trileptons} and in particular of this class of models
\cite{LNWZ}.  The trilepton rates in the no-scale model have been given in Ref.
\cite{LNWZ}; in Fig. \ref{Figure6} we show these for the case $m_t=130\GeV$.
One can show that with ${\cal L}=100\ipb$ of integrated luminosity, chargino
masses as high as $\approx175\GeV$ could be explored, although some regions of
parameter space for lighter chargino masses would remain unexplored. We expect
that somewhat weaker results will hold for the dilaton model, since the
sparticle masses are heavier in that model, especially the sleptons which
enhance the leptonic branching ratios when they are light enough.
\item (c) The relation $m_{\tilde q}\approx m_{\tilde g}$ for the $\tilde
u_{L,R},\tilde d_{L,R}$ squark masses should allow to probe the low end of the
squark and gluino allowed mass ranges, although the outlook is more promising
for the dilaton model since the allowed range starts off at lower values of
$m_{\tilde g,\tilde q}$ (see Eq. (\ref{gmin})). An important point
distinguishing the two models is that the average squark mass is slightly below
(above) the gluino mass in the no-scale (dilaton) model, which should have an
important bearing on the experimental signatures and rates \cite{sgdetection}.
In the dilaton case the $\tilde t_1$ mass can be below $100\GeV$ for
sufficiently low $m_t$, and thus may be detectable. As the lower bound on $m_t$
rises, this signal becomes less accessible. The actual reach of the Tevatron
for the above processes depends on its ultimate integrated luminosity.
\end{description}

\begin{figure}[t]
\vspace{4.8in}
\vspace{-2.2in}
\caption{\baselineskip=12pt
The number of trilepton events at the Tevatron per $100\ipb$ in the no-scale
model for $m_t=130\GeV$. Note that with $200\ipb$ and 60\% detection efficiency
it should be possible to probe chargino masses as high as $175\GeV$.}
\label{Figure6}
\end{figure}

\subsection{LEPI,II}
\begin{description}
\item (a) In the class of models we consider, the lightest Higgs boson has
couplings to gauge bosons and fermions which are close to those of the Standard
Model (SM) Higgs boson, and therefore experimental lower bounds to the SM Higgs
mass have been shown to apply slightly weakened to the supersymmetric Higgs
\cite{LNPWZh}. Since the lower bound on the SM Higgs boson mass could still be
pushed up several GeV at LEPI, the strict dilaton scenario (which requires
$m_h\approx61-91\GeV$) could be further constrained at LEPI and definitely
tested at LEPII. At LEPII the SM Higgs mass could be explored up to roughly the
beam energy minus $100\GeV$ \cite{Alcaraz}. This will allow exploration of the
low $\tan\beta$ values in both models, although the strict no-scale case will
probably be out of reach (see Figs. \ref{Figure4},\ref{Figure5}). The
$e^+e^-\to hA$ channel will be open for large $\tan\beta$ and low $m_{\tilde
g}$.
\item (b) Chargino masses below the kinematical limit
($m_{\chi^\pm_1}\lsim100\GeV$) should not be a problem to detect through the
``mixed" mode with one chargino decaying leptonically and the other one
hadronically \cite{LNPWZ}, \ie, $e^+e^-\to\chi^+_1\chi^-_1$, $\chi^+_1\to
\chi^0_1 q\bar q'$, $\chi^-_1\to\chi^0_1 l^-\bar\nu_l$. In Fig. \ref{Figure7}
(top row) we show the correponding event rates in the no-scale model. Note that
$m_{\chi^\pm_1}$ can be as high as $\approx290\GeV$ in these models.
\item (c) Selectron, smuon, and stau pair production is partially accessible
for both the no-scale and dilaton models, although more so in the no-scale
case. In Fig. \ref{Figure7} (bottom row) we show the rates for the most
promising (dilepton) mode in $e^+e^-\to\tilde e^+_R\,\tilde e^-_R$ production
in the no-scale model.
\end{description}

\begin{figure}[t]
\vspace{4.7in}
\vspace{-0.3in}
\caption{\baselineskip=12pt
The number of ``mixed" events (1-lepton+2jets+$\mpt$) events per ${\cal
L}=100\ipb$ at LEPII versus the chargino mass in the no-scale model (top row).
Also shown (bottom row) are the number of di-electron events per ${\cal
L}=100\ipb$  from selectron pair production versus the lightest selectron
mass.}
\label{Figure7}
\end{figure}

\subsection{HERA} The elastic and deep-inelastic contributions to
$e^-p\to\tilde e^-_R\chi^0_1$ and $e^-p\to\tilde\nu\chi^-_1$ in the no-scale
model should push the LEPI lower bounds on the lightest selectron, the lightest
neutralino, and the sneutrino masses by $\approx25\GeV$ with ${\cal L}=100\ipb$
\cite{hera}. In Fig. \ref{Figure8} we show the elastic plus deep-inelastic
contributions to the total supersymmetric signal ($ep\to{\rm susy}\to eX+\mpt$)
versus the lightest selectron mass ($m_{\tilde e_R}$) and the sneutrino mass
$(m_{\tilde\nu})$ in the no-scale model. These figures show the ``reach" of
HERA in each of these variables. With ${\cal L}=1000\ipb$ HERA should be
competitive with LEPII as far as the no-scale model is concerned. In the
dilaton scenario, because of the somewhat heavier sparticle masses, the
effectiveness of HERA is reduced, although probably both channels may be
accessible.

\begin{figure}[t]
\vspace{4.7in}
\vspace{-0.3in}
\caption{\baselineskip=12pt
The elastic plus deep-inelastic total supersymmetric cross section at HERA
($ep\to{\rm susy}\to eX+\mpt$) versus the lightest selectron mass ($m_{\tilde
e_R}$) and the sneutrino mass ($m_{\tilde\nu}$). The short- and long-term
limits of sensitivity are expected to be $10^{-2}\pb$ and $10^{-3}\pb$
respectively.}
\label{Figure8}
\end{figure}

\section{Conclusions}
We have presented the simplest, string-derivable, supergravity model which
has as gauge group flipped $SU(5)$ with supplementary matter representations
to ensure unification at the string scale ($\sim10^{18}\GeV$). This basic
structure is complemented by two possible string supersymmetry breaking
scenaria: $SU(N,1)$ no-scale supergravity and dilaton-induced supersymmetry
breaking. These two variants should be considered to be idealizations of what
their string-derived incarnation should be. The specification of the hidden
sector is crucial to the determination of the supersymmetry breaking
scenario at work. A thorough exploration of the parameter spaces of the
two models yields interesting results for experimental detection at present
or near future colliders. In this regard, the no-scale model is more within
reach than the dilaton model, because of its generally lighter spectrum.
In both supersymmetry breaking scenaria considered, there ia a more constrained
special case which allows $\tan\beta$ to be determined in terms of $m_t$ and
$m_{\tilde g}$. In the strict no-scale case we find a striking result: if
$\mu>0$, $m_t\lsim135\GeV$, whereas if $\mu<0$, $m_t\gsim140\GeV$. Therefore
the value of $m_t$ determines the sign of $\mu$. Furthermore, we found that the
value of $m_t$ also determines  whether the lightest Higgs boson is above or
below $100\GeV$. In the restricted dilaton case there is an upper bound on
the top-quark mass ($m_t\lsim155\GeV$) and the lightest Higgs boson mass
($m_h\lsim91\GeV$). Thus, continuing Tevatron top-quark searches and LEPI,II
Higgs searches could probe this restricted scenario completely. In Table
\ref{Table4} we give a summary of the general properties of these models and
a comparison of their spectra.

\begin{table}[p]
\hrule
\caption{Major features of the $SU(5)\times U(1)$ string-inspired/derived model
and a comparison of the two supersymmetry breaking scenaria considered.
(All masses in GeV).}
\label{Table4}
\begin{center}
\begin{tabular}{|l|}\hline
\hfil$\bf SU(5)\times U(1)$\hfil\\ \hline
$\bullet$ Easily string-derivable, several known examples\\
$\bullet$ Symmetry breaking to Standard Model due to vevs of \r{10},\rb{10}\\
\quad and tied to onset of supersymmetry breaking\\
$\bullet$ Natural doublet-triplet splitting mechanism\\
$\bullet$ Proton decay: $d=5$ operators very small\\
$\bullet$ Baryon asymmetry through lepton number asymmetry\\
\quad (induced by the decay of flipped neutrinos) as recycled by\\
\quad non-perturbative electroweak interactions\\
\hline
\end{tabular}
\end{center}
\begin{center}
\begin{tabular}{|l|l|}\hline
\hfil$\vev{F_M}_{m_0=0}$ (no-scale)\hfil&\hfil$\vev{F_D}$ (dilaton)\hfil\\
									\hline
$\bullet$ Parameters 3: $m_{1/2},\tan\beta,m_t$&
$\bullet$ Parameters 3: $m_{1/2},\tan\beta,m_t$\\
$\bullet$ Universal soft-supersymmetry&$\bullet$ Universal soft-supersymmetry\\
\quad breaking automatic&\quad breaking automatic\\
$\bullet$ $m_0=0$, $A=0$&$\bullet$ $m_0=\coeff{1}{\sqrt{3}}m_{1/2}$,
$A=-m_{1/2}$\\
$\bullet$ Dark matter: $\Omega_\chi h^2_0<0.25$
&$\bullet$ Dark matter: $\Omega_\chi h^2_0<0.90$\\
$\bullet$ $m_{1/2}<475\GeV$, $\tan\beta<32$
&$\bullet$ $m_{1/2}<465\GeV$, $\tan\beta<46$\\
$\bullet$ $m_{\tilde g}>245\GeV$, $m_{\tilde q}>240\GeV$
&$\bullet$ $m_{\tilde g}>195\GeV$, $m_{\tilde q}>195\GeV$\\
$\bullet$ $m_{\tilde q}\approx0.97m_{\tilde g}$
&$\bullet$ $m_{\tilde q}\approx1.01m_{\tilde g}$\\
$\bullet$ $m_{\tilde t_1}>155\GeV$
&$\bullet$ $m_{\tilde t_1}>90\GeV$\\
$\bullet$ $m_{\tilde e_R}\approx0.18m_{\tilde g}$,
$m_{\tilde e_L}\approx0.30m_{\tilde g}$
&$\bullet$ $m_{\tilde e_R}\approx0.33m_{\tilde g}$,
$m_{\tilde e_L}\approx0.41m_{\tilde g}$\\
\quad $m_{\tilde e_R}/m_{\tilde e_L}\approx0.61$
&\quad $m_{\tilde e_R}/m_{\tilde e_L}\approx0.81$\\
$\bullet$ $60\GeV<m_h<125\GeV$&$\bullet$ $60\GeV<m_h<125\GeV$\\
$\bullet$ $2m_{\chi^0_1}\approx m_{\chi^0_2}\approx m_{\chi^\pm_1}\approx
0.28 m_{\tilde g}\lsim290$
&$\bullet$ $2m_{\chi^0_1}\approx m_{\chi^0_2}\approx m_{\chi^\pm_1}\approx
0.28 m_{\tilde g}\lsim285$\\
$\bullet$ $m_{\chi^0_3}\sim m_{\chi^0_4}\sim m_{\chi^\pm_2}\sim\vert\mu\vert$
&$\bullet$ $m_{\chi^0_3}\sim m_{\chi^0_4}\sim m_{\chi^\pm_2}\sim\vert\mu\vert$
\\
$\bullet$ Spectrum easily accessible soon
&$\bullet$ Spectrum accessible soon\\ \hline
$\bullet$ Strict no-scale: $B(M_U)=0$
&$\bullet$ Special dilaton: $B(M_U)=2m_0$\\
\quad $\tan\beta=\tan\beta(m_t,m_{\tilde g})$
&\quad $\tan\beta=\tan\beta(m_t,m_{\tilde g})$\\
\quad $m_t\lsim135\GeV\Rightarrow\mu>0,m_h\lsim100\GeV$
&\quad $\tan\beta\approx1.4-1.6$, $m_t<155\GeV$\\
\quad $m_t\gsim140\GeV\Rightarrow\mu<0,m_h\gsim100\GeV$
&\quad $m_h\approx61-91\GeV$\\
\hline
\end{tabular}
\end{center}
\hrule
\end{table}

We conclude that these well motivated string-inspired/derived models
(especially their strict versions) could soon be probed experimentally. The
various ingredients making up these models are likely to be present
in actual fully string-derived models which yield the set of supersymmetry
breaking parameters in Eqs. (\ref{nsc},\ref{kl}). The search for such models is
imperative, although it may not be an easy task since in traditional gaugino
condensation scenaria Eqs. (\ref{nsc},\ref{kl}) are usually not reproduced
(see however Refs. \cite{Ross,Zwirner}). Moreover, the requirement of vanishing
vacuum energy may be difficult to fulfill, as a model with these properties and
all the other ones outlined in Sec. 1 is yet to be found. This should not be
taken as a discouragement since the harder it is to find the correct model,
the more likely it is to be in some sense unique.

\section*{Acknowledgements}
This work has been supported in part by DOE grant DE-FG05-91-ER-40633. The work
of J.L. has been supported by an SSC Fellowship.

\end{document}